\documentclass[12pt,a4paper]{article}
\usepackage{a4wide}
\usepackage{graphicx}
\usepackage{subfigure}
\usepackage[centertags]{amsmath}
\usepackage{amssymb}
\usepackage{cite}
\usepackage{epsfig}
\usepackage{longtable}
\usepackage{enumerate}
\usepackage{scrtime}

\newcommand{\id}{1\!\!1}

\begin{document}

\thispagestyle{empty}

\begin{flushright}
                     IPPP/06/57\\
                     DCPT/06/114\\
\end{flushright}
\vskip 2cm

\begin{center}
{\huge Gauging Flavour in Meta-Stable Susy Breaking Models }
\end{center}
\vspace*{5mm} \noindent
\vskip 0.5cm
\centerline{\bf Stefan F\"orste}

\vskip 1cm

\centerline{\em Institute for Particle Physics Phenomenology (IPPP)}
\centerline{\em South Road, Durham DH1 3LE, United Kingdom}
\vskip2cm

\centerline{\bf Abstract}
\vskip .3cm

We modify the first ISS model \cite{Intriligator:2006dd} by
gauging a diagonal flavour symmetry. We add additional multiplets
transforming as fundamentals and anti-fundamentals under the gauged
flavour group. Their number is chosen such that the microscopic theory
is asymptotically free whereas in the Seiberg dual (w.r.t.\ the colour
group) it changes to an infrared free theory. Non
perturbative effects within the flavour group can correct the location
of the supersymmetric vacuum.  
Statements about meta-stability of the susy breaking vacuum would
require a two loop calculation. For general couplings, the question
whether gauging 
flavour destabilises susy breaking remains open.  

\vskip .3cm

\newpage

\section{Introduction}

The idea of meta-stable susy breaking \cite{Dimopoulos:1997ww} has
recently been realised in appealingly simple models
\cite{Intriligator:2006dd}. The underlying microscopic theory can be
as simple as supersymmetric QCD with a quadratic superpotential. There
is a point in the field space of the theory where a dual macroscopic
description is in its perturbative regime and only susy breaking vacua
exist. The susy breaking scale is a function of a dynamically
generated scale and a tree level mass. At another point in moduli
space the macroscopic theory is strongly coupled at the susy breaking
scale. Due to non perturbative corrections a supersymmetric vacuum
exists at that point. The tunnelling from the susy breaking to the susy
preserving vacuum is, however, highly suppressed and the mechanism is
stable enough for all practical purposes. The authors of
\cite{Intriligator:2006dd} indicate that they would like to use their
mechanism for direct mediation of susy breaking. 

In principle, one could also try to use it for breaking susy in a hidden
sector and mediate the breaking to the visible sector by one of the
known mechanisms. Indeed, such constructions have been put forward in 
\cite{Garcia-Etxebarria:2006rw}. In the present note, we investigate
as a toy model for this possibility a simple modification of
\cite{Intriligator:2006dd}, where the flavour group is gauged. In
order to avoid strong gauge couplings at the susy breaking point in
field space we have to add some colour singlets transforming non
trivially under flavour. In the non super symmetric vacuum there are
fields which remain massless after one loop corrections. 

The moduli space of supersymmetric vacua
within this theory is not completely understood. With a certain amount
of fine tuning on the flavour coupling we can keep non perturbative
corrections due to gauged flavour small at the susy breaking scale and
compute first order corrections of such effects to the location of the
supersymmetric vacuum. The first order correction can move the
supersymmetric vacuum towards the non supersymmetric one. 
Our calculation is only valid if these corrections are small. This is
indeed the case and hence the calculation is self consistent. 

In the next section we review the model
\cite{Intriligator:2006dd}.  
In section three we gauge the flavour symmetry and add colour singlets
transforming as fundamentals and anti-fundamentals under the flavour
group. We complete the electric-magnetic dictionary for this case. 
In section four we discuss meta-stable susy breaking for the modified
model. There are mass squareds which could become negative at a two
loop level. We compute corrections to the location of the
supersymmetric vacuum due to non perturbative effects in the gauged
flavour sector, in an approximation where these corrections are necessarily
small. Finally, we provide some concluding remarks.  
In spirit, the discussed model is similar to \cite{Ooguri:2006pj}
where gauged flavour has been considered in the context of quiver
theories.

\section{The ISS model}

Several models for meta-stable supersymmetry breaking have recently been
presented in  \cite{Intriligator:2006dd}, and subsequently in
\cite{Ooguri:2006pj,Franco:2006es,Garcia-Etxebarria:2006rw,Kitano:2006wm,
  Braun:2006em,
  Argurio:2006ew,Braun:2006da,Banks:2006ma,Ray:2006wk}. 
Here,  
the focus will be on the first model in \cite{Intriligator:2006dd} which we
briefly review in the following. 
The model has two dual descriptions: One, the
microscopic model, 
is appropriate for calculations in the ultraviolet region, since it is
asymptotically free, and has no further non renormalisable
couplings. The dual description, or the macroscopic model, is 
infrared free, and hence 
the adequate choice for investigating phenomena at lower energy. In
the macroscopic model,
supersymmetry breaking occurs at lower energies, where the
perturbative description is reliable. 

The microscopic model is an $SU(N_c)$ gauge theory with $N_f$
fundamentals, combined into an $N_f \times N_c$ matrix $Q$,  and
anti-fundamentals, written into an $N_c \times N_f$ matrix
$\tilde{Q}$, with
\begin{equation} \label{eq:region}
N_c + 1 \leq N_f < \frac{3}{2} N_c .
\end{equation}
There is a tree level potential\footnote{We do not introduce a general 
  mass matrix breaking this symmetry completely, since later we want to
  consider a model where the unbroken diagonal subgroup is gauged.} 
\begin{equation} \label{eq:upsup}
W= m \mbox{Tr} \left( Q \tilde{Q}\right)
\end{equation}
breaking the global $SU(N_f)\times SU(N_f)$ flavour symmetry to its
diagonal subgroup. 

The Seiberg dual \cite{Seiberg:1994pq} (for reviews see e.g.\
\cite{Intriligator:1995au,Giveon:1996qs,Berman:2002kd}) of this
theory has gauge group $SU(N)$ with 
\begin{equation}
N = N_f - N_c,
\end{equation}
$N_f$ quarks $\varphi$, $N_f$ anti-quarks $\tilde{\varphi}$ and singlets
$\Phi$ transforming as an adjoint plus singlet under the diagonal
flavour.  The superpotential is  
\begin{equation}\label{eq:downsup}
W = h \mbox{Tr}\left( \tilde{\varphi} \Phi \varphi\right) - h \mu^2
\mbox{Tr}\Phi .
\end{equation}
The dictionary for the couplings and parameters reads
\cite{Intriligator:2006dd}
\begin{eqnarray}
\hat{\Lambda}^{N_f} & = &
 \Lambda_c ^{3N_c - N_f} \tilde{\Lambda}_c ^{3N- N_f}\left(
 -1\right)^N,\label{eq:hatlam}  \\ 
\varphi = q,\,\,\, \tilde{\varphi} = \tilde{q}, & & h =
 \frac{\sqrt{\alpha}\Lambda_c}{\hat{\Lambda}}, \,\,\, \mu^2 = - m
\hat{\Lambda} , \\ \Phi  & = & 
\frac{Q\tilde{Q}}{\sqrt{\alpha} \Lambda_c} ,\label{eq:rescale} 
\end{eqnarray}
where $\Lambda_c$ and $\tilde{\Lambda}_c$ are the scales at which the
colour couplings of the microscopic and macroscopic theory diverge,
respectively. The scale $\hat{\Lambda}$ is an additional scale
appearing in the dual theory.
$Q$, $\tilde{Q}$ are related to $q$, $\tilde{q}$ by Hodge
duality on the baryons as in \cite{Intriligator:1995au}.  
The factor in (\ref{eq:rescale}) appears due to a
rescaling giving $\Phi$ mass dimension one and removing an order one
factor $\alpha$ from the K\"ahler potential
\cite{Intriligator:2006dd}.
This will be of some importance in the next section.

The major observation of \cite{Intriligator:2006dd} is the existence
of a meta-stable non supersymmetric vacuum in the perturbative regime
of the macroscopic model, 
\begin{equation}\label{eq:nonsumin}
\Phi = 0 \,\,\, ,\,\,\, \varphi^T = \tilde{\varphi} = \left( \id_N ,
0_{N_f -N}\right) . 
\end{equation}
In this vacuum, some fields have tree level masses of the order
$|h\mu|$. In particular $\mbox{Tr}_N \Phi$ is classically massive,
where the index $N$ indicates that the trace is taken over the first
$N$ diagonal entries. Other fields, the pseudo moduli, become massive
at one loop of perturbation theory. The Goldstone bosons of broken
global symmetries remain massless.

In the supersymmetric vacuum, all matter fields are massive and 
$\Phi$ has the expectation value
\begin{equation}\label{eq:susol}
\Phi = \frac{\tilde{\Lambda}_c}{h} \left(
\frac{\mu}{\tilde{\Lambda}_c} \right)^{\frac{2N}{N_f -N}}\id_{N_f} .
\end{equation}
Translated to the microscopic theory,
this corresponds to
\begin{equation} \label{eq:susyvac}
Q\tilde{Q} = m^{2}\left( \frac{\Lambda_c}{m}
\right)^{\frac{3N_c - N_f}{N_c}}\id_{N_f} .
\end{equation}
At the supersymmetry breaking vacuum this VEV is zero. So, if the quark masses
are well below the strong coupling scale of the microscopic theory,
the susy breaking vacuum and the supersymmetric vacuum are well
separated. This way, the tunnelling probability to the
supersymmetric vacuum can be suppressed parametrically.  

\section{Dualising Colour with Gauged Flavour Watching}

We modify the microscopic model of the previous section by gauging the
diagonal flavour group $SU(N_f)$ and add colour singlets transforming
under the fundamental and anti-fundamental representation of $SU(N_f)$.
We add $k$ of such pairs, with
\begin{equation}\label{eq:krange}
N_f + N_c < k < 3N_f -N_c.
\end{equation}
(Note that $k > 3/2 N_f$ is implied by (\ref{eq:region}).)
So, altogether we have gauge group 
\begin{equation}
SU\left( N_c \right) \times SU\left( N_f\right) ,
\end{equation} 
and 
\begin{itemize}
\item one chiral multiplet, $Q$, in the $\left( N_c ,
  \overline{N}_f\right)$, 
\item one chiral multiplet, $\tilde{Q}$, in the $\left( \overline{N}_c ,
  N_f\right)$ ,
\item $k$ chiral multiplets, $\rho$, in the $\left( 1,
  N_f\right)$,
\item $k$ chiral multiplets, $\tilde{\rho}$, in the $\left(
  1, \overline{N}_f\right)$ .
\end{itemize}
The one loop beta function coefficient of the flavour coupling
is greater than zero, and hence the microscopic theory is well defined
in the UV region. For the tree level potential we take 
\begin{equation}
W = m \mbox{Tr} \left( Q \tilde{Q}\right) + \sigma
\mbox{Tr}\left(\rho\tilde{\rho} \right) ,
\end{equation}
where the mass sigma is taken to be far below the susy breaking scale,
in order to ensure that these degrees of freedom contribute to the
running coupling at and above the susy breaking scale. 
We assume\footnote{Here, and in the following, it should be understood
  that we compare real parts of
  dynamically generated scales.}
\begin{equation} \label{eq:flasma}
\Lambda_f < \Lambda_c ,
\end{equation}
where $\Lambda_f$ is the scale at which the flavour coupling
diverges. In the region where the flavour coupling is still small
whereas the colour coupling becomes large, we replace the theory by an
equivalent formulation. The dual theory is obtained by performing the
same duality as in the previous section on the colour group. 
One obtains a theory with
dual gauge group
\begin{equation}
SU(N=N_f - N_c) \times SU(N_f) .
\end{equation}
The dual spectrum consists of
\begin{itemize}
\item one chiral multiplet, $\varphi$, in the $\left( N ,
  \overline{N}_f\right)$, 
\item one chiral multiplet, $\tilde{\varphi}$, in the $\left( \overline{N} ,
  N_f\right)$ ,
\item $k$ chiral multiplets, $\rho$, in the $\left( 1,
  N_f\right)$,
\item $k$ chiral multiplets, $\tilde{\rho}$, in the $\left(
  1, \overline{N}_f\right)$ ,
\item one chiral multiplet $\Phi$ in the $(1,1) + \left(1, N_f^2 -1\right)$.
\end{itemize}
As before, the one loop beta function coefficient of the colour group
is negative. If $k$ satisfies (\ref{eq:krange}), the one loop beta function
coefficient of the flavour group is less than zero. 
Hence, the flavour group
stays in its perturbative regime as long as none of the flavour
charged matter becomes massive. 

Next, we need to identify the scale
$\tilde{\Lambda}_f$,  at which the dual flavour coupling becomes
large. We propose that the relation is
\begin{equation}\label{eq:dualfla}
\Lambda_f ^{3N_f - N_c -k}\tilde{\Lambda}_f ^{-\left(3N_f -N_f -N -
  k\right)} = 
\hat{\Lambda}^{-N} \Lambda_c^{3\left( N_f -
  N_c\right)}\alpha^{N_f}2^{- N_f} .
\end{equation}
We have no strict derivation for (\ref{eq:dualfla}). We can provide
only some reasoning. The $\hat{\Lambda}$ dependence is fixed by
imposing that 
$\hat{\Lambda}$ parameterises unknown factors in the K\"ahler
potential for $\varphi$ and $\tilde{\varphi}$
\cite{Intriligator:2006dd}. 
In the theory with ungauged flavour, discussed in the previous
section, there is a scaling symmetry of the superpotential and the
matching relation (\ref{eq:hatlam}) involving dual quantities
\begin{equation}\label{eq:symmetry}
\varphi \to \gamma \varphi , \,\,\, \tilde{\varphi} \to \gamma
\tilde{\varphi}, \,\,\, \hat{\Lambda} \to \gamma^2
\hat{\Lambda},\,\,\, \Phi \to \gamma^{0}\Phi .
\end{equation}
Our proposal (\ref{eq:dualfla}) respects that symmetry.
Since $\Lambda_c$ is used throughout, in order to obtain canonical mass
dimensions, we did so in (\ref{eq:dualfla}) as well\footnote{Taking a
  flavour scale instead would mean that the relation alters its form
  upon giving one of the pairs in the $\left( 1, N_f\right)$ and $\left( 1,
  \overline{N}_f\right)$ a large mass and integrating it out. Using
  $\tilde{\Lambda}_c$ would spoil the `symmetry' (\ref{eq:symmetry})
  unless it comes together with $\hat{\Lambda}$ in an invariant
  combination. But such a combination is $\Lambda_c$.}. The
$\alpha$ dependence is due to (\ref{eq:rescale}), and the anomalous
rescaling of the adjoint. The numerical coefficient $2^{-N_f}$ will be
discussed in the end of the fourth section.

For the dual tree level potential\footnote{Note that a term
  $\mbox{Tr} 
  \tilde{\rho} \Phi \rho $ would correspond to a non renormalisable
coupling in the microscopic theory. We do not consider this case.} we
obtain
\begin{equation}\label{eq:downsupg}
W = h \mbox{Tr}\left( \tilde{\varphi} \Phi \varphi\right) - h \mu^2
\mbox{Tr}\Phi  + \sigma
\mbox{Tr}\left(\rho\tilde{\rho}\right).
\end{equation}

In the dual theory we are interested in a situation where the dual
flavour coupling diverges farther in the ultraviolet than the dual colour
coupling
\begin{equation}\label{eq:inequ}
\tilde{\Lambda}_c < \tilde{\Lambda}_f .
\end{equation}
This condition is not invariant under the previously discussed scaling
symmetry. One has to supplement it with the requirement of canonical
kinetic terms. 

To see whether this involves some fine tuning we combine
(\ref{eq:hatlam}) and (\ref{eq:dualfla}) into 
\begin{equation}
\Lambda_f ^{2\left(N_f-N_c\right)} \left(
  \frac{\tilde{\Lambda}_f}{\Lambda_f}\right) 
  ^{k +N_f+N - 3N_f} = \left(\frac{\alpha}{2}\right)^{N_f}\left(
  \Lambda_c\right)^{2\left( 
  N_f-N_c\right)} \left( 
  \frac{\tilde{\Lambda}_c}{\Lambda_c}\right)^{\frac{N\left(N_f
  -3N\right)}{N_f}}.  
\end{equation}
The hierarchy imposed by $\Lambda_f < \Lambda_c$ and
$\tilde{\Lambda}_f > \tilde{\Lambda}_c$ implies  
\begin{equation}\label{eq:fine}
\left(\frac{\alpha}{2}\right)^{N_f}\left( \frac{
  \Lambda_c}{\Lambda_f}\right)^{2\left(N_f-N_c\right)} \left( 
\frac{\tilde{\Lambda}_c}{\Lambda_c}\right)^{\tilde{b}_f -
  \frac{N}{N_f} \tilde{b}_c} > 1 ,
\end{equation}
with
\begin{equation}
\tilde{b}_f = 3N_f -N - N_f - k ,\hspace*{2em} \tilde{b}_c = 3N - N_f ,
\end{equation}
being the (negative) one loop beta function coefficients of dual
flavour and colour, respectively.
We require further that there is a region where electric and magnetic
gauge couplings are finite, i.e.\  $\tilde{\Lambda}_c /\Lambda_c >1$. 
The inequality (\ref{eq:fine}) holds naturally, if the dual colour
coupling diverges fast enough, compared to the flavour coupling, for the
exponent in (\ref{eq:fine}) to be positive. Otherwise some fine tuning on
$\Lambda_c/\Lambda_f$ is needed. 

\section{Meta-Stable Susy Breaking with Gauged Flavour?}

As in \cite{Intriligator:2006dd} there is no supersymmetric vacuum in
the perturbative regime.
The potential is minimised by (\ref{eq:nonsumin}). The colour group is
broken completely whereas the flavour group is broken to $SU\left(N_f
- N = N_c\right)$. Some of the classical moduli provide the missing
degrees of freedom for massive vector multiplets of broken
flavour. Others acquire a D-term tree level mass due to the equations
of motion of the flavour gauge degrees of freedom. 
Our condition that $\sigma$ is smaller than the susy breaking scale
means that at the susy breaking scale quantum corrections to the mass
of $\rho$ and $\tilde{\rho}$ dominate over the tree level mass
$\sigma $. Hence, $\rho$ and $\tilde{\rho}$ can
effectively considered to be massless.  
One loop quantum corrections to their mass 
cancel since supersymmetry breaking is mediated only by gauge degrees
of freedom whose mass spectrum is classically fermion-boson
degenerate \cite{Intriligator:2006dd}. If they are not stabilised by
higher order quantum 
corrections they can gain a vacuum expectation value up to
$\sigma\mbox{Tr}\left< \tilde{\rho}\rho\right> $ being
of the order of the susy breaking scale. 
The situation is similar to one of the examples in
\cite{Franco:2006es} where it was concluded that the given information
is not sufficient for making statements about the stability of the
vacuum. Instead of speculating about the nature of higher order
quantum corrections to the susy breaking vacuum, we try to
get insight into non perturbative flavour corrections to the
supersymmetric vacuum.

To this end, we give a VEV to $\Phi$. Then the bifundamentals
$\varphi$, $\tilde{\varphi}$ become massive. Below this mass both gauge groups
become asymptotically free, if $k < 2 N_f$. (As we will see shortly
the adjoint becomes also massive and below that mass the flavour group
is always asymptotically free.)

The supersymmetric moduli
space of the colour sector is well understood
\cite{Seiberg:1994bz,Intriligator:1995au}. Due to the presence of the
adjoint (traceless part of $\Phi$) the flavour sector is more
complicated. There are some partial results in the literature
\cite{Kutasov:1995ve,Kutasov:1995np,Kutasov:1995ss,Elitzur:1995ap,
  Elitzur:1995xp,Hanany:1995na}. It is, however, not obvious to us how
to use these 
results in the given situation. Therefore, we will impose the
simplifying assumption that the flavour group becomes strongly coupled
at much lower energies than the colour group\footnote{After decoupling the
bifundamentals, the one loop beta function coefficients are:
$3\left(N_f - N_c\right)$ for the magnetic colour group, and at most
$N_f - N_c$ for the 
flavour group. So, the colour coupling will diverge at higher
scales than the flavour one.}. We assume further, that non perturbative
effects within the flavour group can be neglected at the mass of the
bifundamentals where the two theories decouple. 

Due to instanton corrections within the colour group the
superpotential (\ref{eq:downsupg}) is supplemented by an additional
contribution\footnote{For the case $N_f = N_c +1$ see the discussion
  in \cite{Intriligator:2006dd}.} 
\begin{equation}\label{eq:nonperco}
W^{np}_c = N \left( \tilde{\Lambda}_c^{-\left(N_f - 3N\right)} \det
h\Phi \right)^{1/N} . 
\end{equation}
The traceless part of $\Phi$ acquires a mass
\begin{equation}\label{eq:admass}
m_{N_f ^2 -1} = -\frac{h^2}{2N_f} 
\mbox{Tr} h\Phi  
\left( \frac{ \mbox{Tr} h\Phi}{N_f \tilde{\Lambda}_c}\right)^{\frac{N_f -
  3N}{N}} .
\end{equation}
After integrating out the traceless part, we fix $\Phi$ to be of the form
\begin{equation}\label{eq:quarkmass} 
\Phi = p \id_{N_f} ,
\end{equation}
and the masses of the bifundamentals become
\begin{equation}
m_ \varphi = m_{\tilde{\varphi}} = h p .
\end{equation}
Assuming that the decoupling of the flavour group at $hp$ occurs above
the scale at which the coupling becomes strong, we can treat its non
perturbative effects separately. After the adjoint and the
bifundamentals decouple we obtain $SU(N_f)$ gauge theory with $k$
pairs of quarks and anti-quarks. 
Now, since $3N_f > k > 3/2 N_f$, the Seiberg dual of the flavour
theory is also asympotically free. However, we can still find a non trivial
supersymmetric vacuum by giving a vacuum expectation value to the dual
of $\mbox{Tr}\rho \tilde{\rho}$. For this case, we can adapt the
result for the supersymmetric vacuum from\footnote{There could be also other
  supersymmetric vacua, we are interested in one giving an effect to
  our considerations.} 
(\ref{eq:susyvac}) 
\begin{equation}\label{eq:solrho}
\rho\tilde{\rho} = \sigma^{2}\left(
\frac{\tilde{\Lambda}_{f,L} }{\sigma}\right)^{\frac{3N_f
      -k}{N_f}}\id_k ,
\end{equation}
where 
\begin{equation}
\tilde{\Lambda}_{L,f}^{3N_f - k} = \tilde{\Lambda}_f ^{ -\left(N +k
  -2N_f\right)} 
\left(hp\right)^{N} m_{N_f ^2 -1} ^{N_f} .
\end{equation}
Plugging this back into (\ref{eq:solrho}) we obtain
\begin{equation}
\sigma \rho\tilde{\rho} = -\frac{1}{2} \sigma^{\frac{k}{N_f}}\left(
h^N \tilde{\Lambda}_f ^{-\left( N+k - 2N_f\right)}\right)
^\frac{1}{N_f} \left( h^{N_f} \tilde{\Lambda}_c^{3N -
  N_f}\right)^\frac{1}{N} p^{\frac{N_f}{N} -\frac{N_f + N_c}{N_f}}\id_k .
\end{equation}
In our approximation the superpotential becomes finally,
\begin{eqnarray}
W & = & N \left( h^{N_f}\tilde{\Lambda}_c^{-\left(N_f - 3N\right)} p^{N_f}
\right)^{1/N}      
-h \mu^2 N_f\, p \nonumber\\ 
& &-\frac{N_f}{2} \sigma^{\frac{k}{N_f}}\left(
h^N \tilde{\Lambda}_f ^{-\left( N+k - 2N_f\right)}\right)
^\frac{1}{N_f} \left( h^{N_f} \tilde{\Lambda}_c^{3N -
  N_f}\right)^\frac{1}{N} p^{\frac{N_f}{N} -\frac{N_f + N_c}{N_f}} ,  
\label{eq:modpot}
\end{eqnarray}
where we have taken into account a contribution from gaugino
condensation to the superpotential (for details see e.g.\
\cite{Intriligator:1995au} section 5.5).

The contribution in the second line of (\ref{eq:modpot}) has to be small
compared to the contribution in the first line, for our approximations
to be consistent. Within this approximation the F-term condition for $\Phi$
can be solved to leading order in the correction,
yielding 
\begin{equation}
\Phi = \Phi_0\left\{ 1 -C\left( 
\sigma^k h^{N }
   \tilde{\Lambda}_f^{-\left(
  k + N - 2N_f\right)}\right)^{\frac{1}{N_f}} \Phi_0 ^{-\frac{N_f +
    N_c}{N_f}} 
\right\} ,
\end{equation}
where $\Phi_0$ is the solution for the ungauged flavour case
(\ref{eq:susol}), and 
\begin{equation}
C = \frac{N}{N_f}\left( \frac{N_f + N_c}{N_f} - \frac{N_f}{N}\right).
\end{equation}
Note, that the correction term contains an $N_f$th root of unity. 

In the microscopic language this corresponds to
\begin{equation}\label{eq:approx}
Q\tilde{Q} = \left(Q\tilde{Q}\right)_0 \left\{ 1 -
2C\sigma^{\frac{k}{N_f} }
\Lambda_f ^{\frac{3N_f - k - N_c}{N_f}} \Lambda_c
  ^{\frac{3N_c-N_f}{N_f}}\left( Q \tilde{Q}\right)_0 ^{-\frac{N_f +
      N_c}{N_f}}\right\} ,
\end{equation}
where $\left(Q\tilde{Q}\right)_0$ denotes the solution in the global
flavour case (\ref{eq:susyvac}).

With our choice of parameters there is
indeed a small correction. If $\sigma$ is taken to zero the
correction vanishes. Our treatment is valid if non
perturbative effects in the flavour sector are negligible at the mass
of the bifundamentals, where magnetic colour and flavour decouple. 

Now, let us come back to the candidate for meta-stable susy breaking,
discussed in the beginning of this section. If the two loop effective
mass squared for $\rho$ and $\tilde{\rho}$ is negative these
fields would condense breaking the residual flavour symmetry. 
Some of the degrees of freedom become part of a massive vector
field. The `mesons' $\rho\tilde{\rho}$ could condense
further. These are singlets under the gauged colour and flavour
groups. 
The backreaction of such a condensation would be
suppressed. Therefore, it seems plausible that the
$\rho\tilde{\rho}$ direction is along a valley where the
position of $\Phi$ and the hight of the potential walls stays effectively
constant. Flavour would
return to a global symmetry which is spontaneously broken in the
meta-stable vacuum. On the supersymmetric side we have seen that the
width of the potential wall can be shortened by the condensation of
$\rho\tilde{\rho}$. Since our approach is limited to a situation
where this shortening of the width is only a small percentage we
cannot say whether such effects could be significant for the life time
of the meta-stable vacuum  in the case that flavour and colour
interactions are of comparable strength at the susy breaking
scale.  

Finally, let us come back to the relation (\ref{eq:dualfla}) and
reason for the numerical coefficient $2^{-N_f}$. Plugging our
approximate solution (\ref{eq:approx}) into (\ref{eq:modpot}) one
obtains
\begin{equation}\label{eq:finsu}
W = N_c \left( \Lambda_c ^{3N_c - N_f} m^{N_f}\right)^{\frac{1}{N_c}}
+ N_f \left( \Lambda_f ^{3N_f - N_c - k}m^{N_c} \sigma^k\right)
^{\frac{1}{N_f}} +\ldots ,
\end{equation}
where dots stand for terms higher order in $\sigma$ which cannot be
computed within our approximation.
The superpotential (\ref{eq:finsu}) is the sum of contributions from gaugino
condensation in the colour and flavour group with the correct
numerical factors (see e.g.\ \cite{Intriligator:1995au}). This
explains our choice in (\ref{eq:dualfla}). 

\section{Conclusions}

Modifying the models of \cite{Intriligator:2006dd} by gauging flavour
can serve as a toy model for gauge mediation of meta-stable susy
breaking, where the hidden sector role is played by the colour gauge
theory  whereas the flavour gauge theory acts as the visible sector. 
We found that in the modified model it is much harder to
derive exact results and make definite statements about
meta-stability. However, we were able to compute corrections to the
location of the supersymmetric vacuum due to non perturbative effects
in the gauged flavour sector as long as these effects are small. 
These corrections shift the position of the supersymmetric vacuum
towards the non supersymmetric one.
Ensuring that non perturbative
effects in the visible sector are highly suppressed at the susy
breaking scale (as well in the supersymmetric vacuum as in the non
supersymmetric one) would keep these corrections small. 
We do not know the first non trivial quantum corrections to some
masses in the non supersymmetric vacuum. This would require a two loop
calculation. Moreover, we lack an exact
knowledge of the moduli space of supersymmetric vacua, for general
couplings. 
In order to improve the situation one could e.g.\ try to modify the
matter sector 
transforming under flavour. For instance, the models discussed in
\cite{Ooguri:2006pj} have also product gauge groups and are under
better control. (However, also the authors of \cite{Ooguri:2006pj}
restrict to a decoupled situation when identifying the supersymmetric
vacuum.) Also recent brane constructions  
\cite{Ooguri:2006bg,Franco:2006ht} may be helpful. 

\bigskip

\noindent {\bf Acknowledgements}

\noindent
I would like to thank David Cerde\~{n}o for useful discussions.


\begin{thebibliography}{51}
%
\bibitem{Intriligator:2006dd}
  K.~Intriligator, N.~Seiberg and D.~Shih,
  JHEP {\bf 0604} (2006) 021
  [arXiv:hep-th/0602239].
%
\bibitem{Dimopoulos:1997ww}
  S.~Dimopoulos, G.~R.~Dvali, R.~Rattazzi and G.~F.~Giudice,
  Nucl.\ Phys.\ B {\bf 510} (1998) 12
  [arXiv:hep-ph/9705307].
%
\bibitem{Garcia-Etxebarria:2006rw}
  I.~Garcia-Etxebarria, F.~Saad and A.~M.~Uranga,
  arXiv:hep-th/0605166.
%
\bibitem{Ooguri:2006pj}
  H.~Ooguri and Y.~Ookouchi,
  arXiv:hep-th/0606061.
%
\bibitem{Franco:2006es}
  S.~Franco and A.~M.~Uranga,
  JHEP {\bf 0606} (2006) 031
  [arXiv:hep-th/0604136].
%
\bibitem{Kitano:2006wm}
  R.~Kitano,
  arXiv:hep-ph/0606129.
%
\bibitem{Braun:2006em}
  V.~Braun, E.~I.~Buchbinder and B.~A.~Ovrut,
  arXiv:hep-th/0606166.
%
\bibitem{Argurio:2006ew}
  R.~Argurio, M.~Bertolini, C.~Closset and S.~Cremonesi,
  arXiv:hep-th/0606175.
%
\bibitem{Braun:2006da}
  V.~Braun, E.~I.~Buchbinder and B.~A.~Ovrut,
  arXiv:hep-th/0606241.
%
\bibitem{Banks:2006ma}
  T.~Banks,
  arXiv:hep-ph/0606313.
%
\bibitem{Ray:2006wk}
  S.~Ray,
  arXiv:hep-th/0607172.
%
\bibitem{Seiberg:1994pq}
  N.~Seiberg,
  Nucl.\ Phys.\ B {\bf 435} (1995) 129
  [arXiv:hep-th/9411149].
%
%
\bibitem{Intriligator:1995au}
  K.~A.~Intriligator and N.~Seiberg,
  Nucl.\ Phys.\ Proc.\ Suppl.\  {\bf 45BC} (1996) 1
  [arXiv:hep-th/9509066].
%
\bibitem{Giveon:1996qs}
  A.~Giveon,
  arXiv:hep-th/9611152.
%
\bibitem{Berman:2002kd}
  D.~S.~Berman and E.~Rabinovici,
  arXiv:hep-th/0210044.
%
\bibitem{Seiberg:1994bz}
  N.~Seiberg,
  Phys.\ Rev.\ D {\bf 49} (1994) 6857
  [arXiv:hep-th/9402044].
%
\bibitem{Kutasov:1995ve}
  D.~Kutasov,
  Phys.\ Lett.\ B {\bf 351} (1995) 230
  [arXiv:hep-th/9503086].
\bibitem{Kutasov:1995np}
  D.~Kutasov and A.~Schwimmer,
  Phys.\ Lett.\ B {\bf 354} (1995) 315
  [arXiv:hep-th/9505004].
%
\bibitem{Kutasov:1995ss}
  D.~Kutasov, A.~Schwimmer and N.~Seiberg,
  Nucl.\ Phys.\ B {\bf 459} (1996) 455
  [arXiv:hep-th/9510222].
%
\bibitem{Elitzur:1995ap}
  S.~Elitzur, A.~Forge, A.~Giveon and E.~Rabinovici,
  Phys.\ Lett.\ B {\bf 353} (1995) 79
  [arXiv:hep-th/9504080].
%
\bibitem{Elitzur:1995xp}
  S.~Elitzur, A.~Forge, A.~Giveon and E.~Rabinovici,
  Nucl.\ Phys.\ B {\bf 459} (1996) 160
  [arXiv:hep-th/9509130].
%
\bibitem{Hanany:1995na}
  A.~Hanany and Y.~Oz,
  Nucl.\ Phys.\ B {\bf 452} (1995) 283
  [arXiv:hep-th/9505075].
%
\bibitem{Ooguri:2006bg}
  H.~Ooguri and Y.~Ookouchi,
  arXiv:hep-th/0607183.
%
\bibitem{Franco:2006ht}
  S.~Franco, I.~Garcia-Etxebarria and A.~M.~Uranga,
  arXiv:hep-th/0607218.
\end{thebibliography}
\end{document}